\documentclass[aps,pre,reprint,groupedaddress,longbibliography]{revtex4-1}

\usepackage{graphicx}
\usepackage{amssymb}
\usepackage{amsmath}
\usepackage{mathtools}
\usepackage{color}
\usepackage[makeroom]{cancel}
\usepackage[colorlinks=true,pdfstartview=Fit]{hyperref} 

\usepackage[normalem]{ulem}		
\usepackage[euler]{textgreek}			
\usepackage{bm}		

\DeclareMathOperator{\erf}{erf}

\begin{document}


\title{When memory pays: Discord in hidden Markov models}


\author{Emma Lathouwers}
\email{elathouw@sfu.ca}
\affiliation{Department of Physics, Simon Fraser University, Burnaby, British Columbia, V5A 1S6, Canada}
\affiliation{Department of Physics, Utrecht University, Princetonplein 5, 3584 CC Utrecht, The Netherlands}

\author{John Bechhoefer}
\email{johnb@sfu.ca}
\affiliation{Department of Physics, Simon Fraser University, Burnaby, British Columbia, V5A 1S6, Canada}


\date{\today}

\begin{abstract}
When is keeping a memory of observations worthwhile?  We use hidden Markov models to look at phase transitions that emerge when comparing state estimates in systems with discrete states and noisy observations.  We infer the underlying state of the hidden Markov models from the observations in two ways: through naive observations, which take into account only the current observation, and through Bayesian filtering, which takes the history of observations into account.  Defining a discord order parameter to distinguish between the different state estimates, we explore hidden Markov models with various numbers of states and symbols and varying transition-matrix symmetry.  All behave similarly.  We calculate analytically the critical point where keeping a memory of observations starts to pay off.  A mapping between hidden Markov models and Ising models gives added insight into the associated phase transitions.
\end{abstract}


\maketitle

\section{Introduction}

Problems requiring statistical inference \cite{mackay03,efron16} are all around us, in fields as varied as neuroscience \cite{hopfield82,coolen05}, signal processing \cite{vanTrees13}, and artificial intelligence (machine learning) \cite{murphy2012machine,goodfellow16}.  A common problem is \textit{state estimation}, where the goal is to learn the underlying state of a dynamical system from noisy observations \cite[Chapt. 10]{murphy2012machine}.  In most cases, the ability to infer states improves smoothly as the signal-to-noise ratio of observations is varied.  However, there can also be phase transitions in the ability to infer the most likely value of a state, as the signal-to-noise ratio of observations is varied \cite{bechhoefer2015hidden}. Formally, phase transitions in inference can occur because problems of inference and statistical physics share common features such as the existence of a free-energy-like function and the requirement or desire that this function be minimized.  Yet the extent to which these elements lead to common outcomes such as phase transitions is not yet clear.

In this paper, we investigate the generality of these links in the context of a specific setting:  the comparison of state estimates based on current observations with those based on both current and past observations.   A simple setting for exploring such problems is given by hidden Markov models (HMMs). They are widely used, from speech recognition \cite{rabiner1989tutorial, huang1990hidden}, to economics \cite{hamilton1989new, mamon2007hidden}, and biology \cite{durbin1998biological, krogh1994hidden}. HMMs describe the evolution of a Markovian variable and the emission of correlated, noisy symbols.  Taking the current emitted symbol at face value gives us a naive state estimate.  However, in these correlated systems there is additional information in the history of emitted symbols, which we can use to find a more refined state estimate.  Comparing the state estimates then reveals in which cases the additional information from keeping a memory of observations makes a difference.

When the observed symbols as a function of time are Markovian, such as HMMs with no noise, there is no advantage to retaining past information. However, for more general systems, the situation is not clear.  Intuitively, if the noise is low (and the entire state vector is observed), then there should be no advantage.  But if the noise is high, then averaging over many observations may help, as long as the system does not change state in the meantime. The surprise is that the transition from a situation where there is no advantage to keeping a memory to one where there is can have the character of a phase transition.  Such transitions have been observed in the specific case of two-state, two-symbol HMMs \cite{bauer14,bechhoefer2015hidden}.

In this article, we ask how general this behavior is in HMMs:  Do we observe these phase transitions \footnote{These transitions have the character of a phase transition in the sense that there is a free-energy-like function that is non-analytic at certain points.  It may not have all characteristics of a thermodynamic phase transition. in more complicated models?} in more complicated models? How sensitive is the behavior of the phase transitions to the details of the model?  And can we understand their origin?  In Section~\ref{sec:theory}, we introduce the theoretical background of the systems we study.  Then, in Sections~\ref{sec:22HMMs}--\ref{sec:2nHMMs}, we introduce and characterize phase transitions in various generalizations of HMMs. In the appendices, we detail the calculation of a phase-transition in $n$-state, $n$-symbol HMMs, and in 2-state, 2-symbol models with broken symmetry.  In an attempt to gain insight into the origins of the observed phase transitions, we also show how to map a two-state, two-symbol HMM onto an Ising model.

\section{State estimation in HMMs}
\label{sec:theory}
  
\begin{figure}[t]
	\centering
	\includegraphics[width=.48\textwidth]{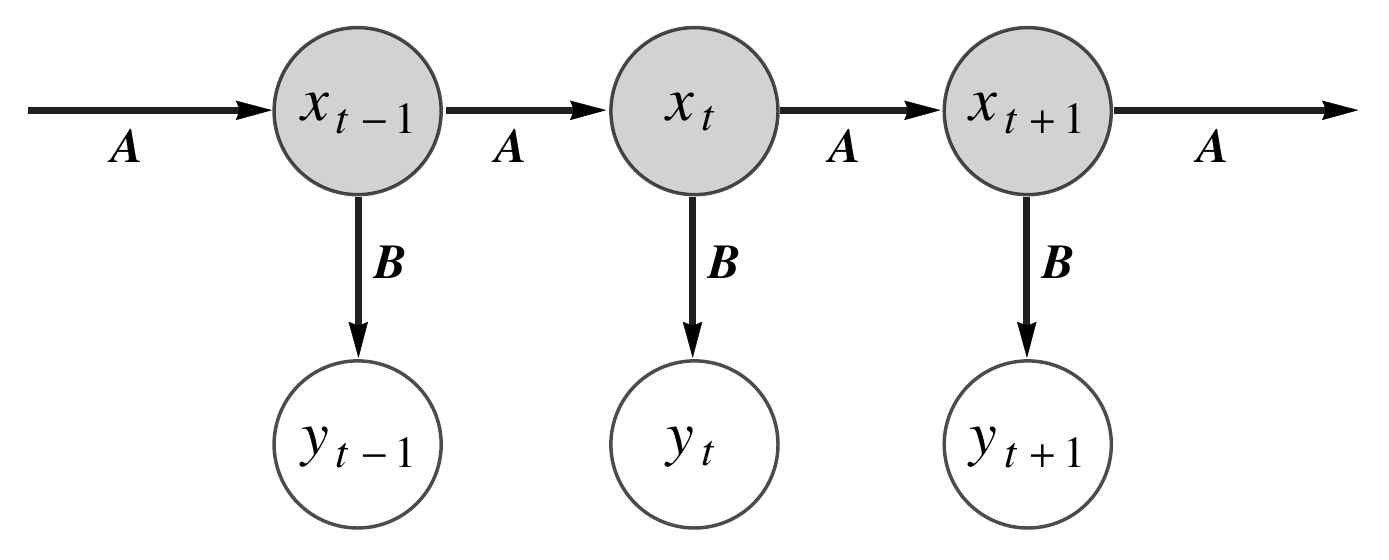}
	\caption{Graphical structure of a HMM.   At time $t$, the hidden state $x_t$ produces an observation $y_t$.  The matrix $\mathbf{A}$ defines the state dynamics, while $\mathbf{B}$ relates observations to states.}
\label{fig:HMM}
\end{figure}

HMMs can be fully described by two probability matrices and an initial state.
The evolution of the hidden state $x_t$ is governed by a $n$-state Markov chain, described by an $n \times n$ transition matrix $\mathbf{A}$ with elements $A_{ij} = P(x_{t+1}=i|x_t=j)$.  The observation of an emitted symbol $y_t$ is described by an $m \times n$ observation matrix $\mathbf{B}$, with elements $B_{ij} = P(y_t=i|x_t=j)$. The matrix dimensions $m$ and $n$ refer to, respectively, the number of symbols and the number of states.  A graphical representation of the dependence structure is shown in Fig.~\ref{fig:HMM}.  The observations depend only on the current state of the system.  Note that the observations as a function of time, described by $P(y_{t+1}|y_t)$, generally do not have Markovian dynamics.  We will refer to an $n$-state, $m$-symbol HMM as an $n \times m$ HMM.

We assume that we have perfect knowledge of our model parameters, and we will focus on comparing state-estimation methods that do or do not keep a memory.  In particular, we will compare the naive observation $y_t$ of the HMM to the state estimate $\hat x_t^{\rm f}$ found through Bayesian filtering.  The Bayesian filtering equations recursively calculate the probability for the system to be in a state $x_t$ given the whole history of observations $y^t$ \cite{bechhoefer2015hidden}, where $y^t = \{ y_1, y_2, \dots, y_t \}$ is used as a shorthand for all past and current information.  The probability is calculated in two steps: the \textit{prediction step} $P(x_{t+1}|y^t)$, and the \textit{update step} $P(x_{t+1}|y^{t+1})$.   The steps can be worked out using marginalization, the definition of conditional probability, the Markov property, and Bayes' theorem.  The transition matrix and the previous filter estimate are needed to predict the next state, and the observation matrix together with the prediction are needed to update the probability.  Together, they give the Bayesian filtering equations \cite{sarkka13},
\begin{subequations}
\label{eq:filter}
	 \begin{align} 
	 \label{eq:filtera}
	  	P(x_{t+1}|y^t)		&= \sum_{x_t} P(x_{t+1}|x_t) P(x_t|y^t) \\[5pt]
	  	P(x_{t+1}|y^{t+1}) 	&= \frac{1}{Z_{t+1}} P(y_{t+1}|x_{t+1}) P(x_{t+1}|y^t) \,,
	\label{eq:filterb}
	\end{align}
\end{subequations}
with the normalization factor 
\begin{align}
	Z_{t+1} &= P(y_{t+1} | y^t) \nonumber \\[5pt]
		&= \sum_{x_{t+1}} P(y_{t+1} | x_{t+1}) P(x_{t+1}|y^t) \,.
\label{eq:normalization}
\end{align}
The Bayesian formulation results in a probability density function for the state $x_t$.  

\begin{figure}[t]
	\centering
	\includegraphics[width=.48\textwidth]{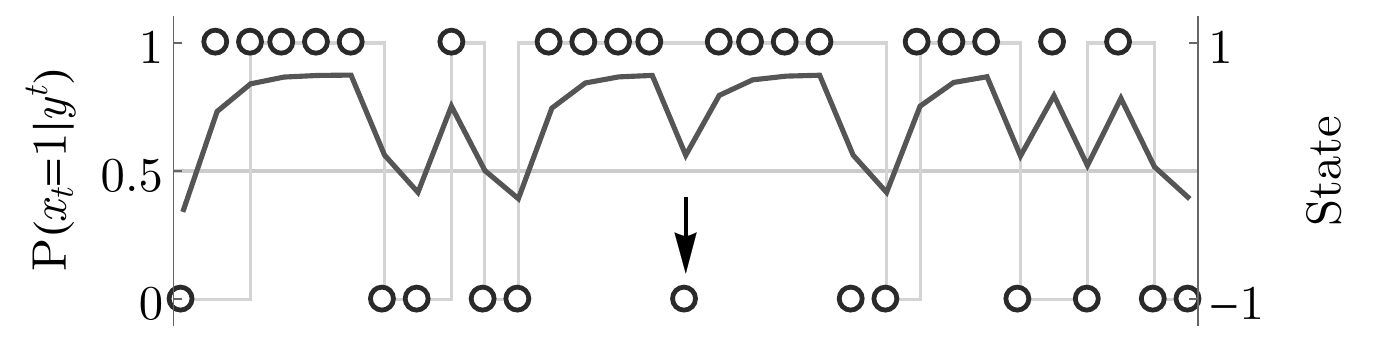}
	\caption{Time series of a $2 \times 2$ HMM generated from a transition matrix $\mathbf{A} = \bigl( \begin{smallmatrix} 0.8 & 0.4 \\ 0.2 & 0.6 \end{smallmatrix} \bigr)$, and an observation matrix $\mathbf{B} = \bigl( \begin{smallmatrix} 0.7 & 0.3 \\ 0.3 & 0.7 \end{smallmatrix} \bigr)$.  The (unknown) Markov chain is shown in light gray, the naive observations as circles, and the filter probability as a black line.  The arrow indicates a time step where the naive state estimate differs from the filter estimate.}
\label{fig:timeseries}
\end{figure}

When the observations $y_t$ are noisy, we cannot be completely sure that our observations and state estimates are correct.  Long sequences of the same observation increase our belief that system is indeed in the observed state, according to Bayesian filtering.  However, even after an infinitely long sequence of the same observation, there is always a chance that the system actually transitioned into another state during the last time step and that we are therefore observing an ``incorrect" symbol (the symbol does not match the state):  The probability to be in state $x_t = i$ given the history of observations $y^t$ is bounded by a maximum confidence level $p^*_i$ that depends on the model's parameters and is defined as the probability to be in a given state after a long sequence of the same observation:
\begin{equation}
	p^*_i = \lim_{t \to \infty} P(x_t = i | y^t = i^t) \, ,
\end{equation}
where by $y^t = i^t$ we mean $\{ y_1 = i, y_2 = i, \dots, y_t = i \}$.
It is important that the sequence of identical observations is long enough that making an additional identical observation does not change the probability. In Fig.~\ref{fig:timeseries} a fragment of the evolution of a HMM, the underlying (unknown) state and observed symbols, is shown together with the Bayesian filtering probability calculated over the time series. For long sequences of identical observations, we see that the filtering probability levels off.
Generally, each state will have a different maximum confidence level. We will return to the maximum confidence level in later calculations and discussions.

Many applications, such as feedback control, depend on single-value estimates $\hat{x}_t$ rather than on probability distributions.  Although statistics such as the mean and median are reasonable candidates for the  ``typical" value of a distribution (minimizing mean-square and absolute errors \cite[Sec.~14.2]{von-der-Linde14}), it is more convenient here to use the mode, or maximum, which is termed, in this context, the \textit{state estimate}.  For state estimates based on all past and current information, we define the \textit{filter estimate}

\begin{equation}
	\hat x_t^{\rm f} \equiv \operatorname*{arg\,max}_{x_t} ( P(x_t|y^t) ) \,.
\label{eq:filter_estimate}
\end{equation}
For the HMM shown in Fig.~\ref{fig:timeseries}, Eq.~\eqref{eq:filter_estimate} implies that whenever the filter probability is above 0.5, the filter estimates the system to be in state 1; similarly, it is in state $-1$ when the probability $<$ 0.5.  Analogously, we define the \textit{naive state estimate} to be based entirely on the current observation, with no use made of past observations,
\begin{equation}
  \hat x_t^{\rm o} \equiv \operatorname*{arg\,max}_{x_t} ( P(x_t|y_t) ) \,.
\end{equation}
In the special case where there is a one-to-one correspondence between symbols $y$ and elements of the internal state $x$, the quantity $\hat{x}_t^{\rm o}$ reduces to $y_t$, the symbol emitted at time $t$.  More generally, the number of internal state components may be smaller than the number of observations, making the interpretation of the estimates more subtle.  As we will see, the combination of defining a probability distribution for the state variable and then selecting its maximum leads to the possibility of phase transitions.

\begin{figure}[t]
	\centering
	\includegraphics[width=0.4\textwidth]{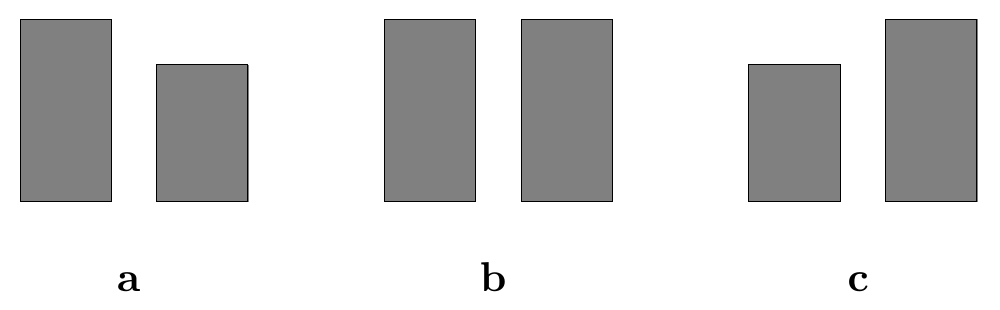}
	\caption{Discrete probability distribution, with continuously varying control parameter going from (a)--(c).}
\label{fig:argmax}
\end{figure}

When one uses other ways to characterize the state than the mode, e.g. the mean, one may not find the analytical discontinuities that we study.  However, the arg max captures an interesting complexity of the probability density function that would be lost in taking the mean.  This is illustrated in Fig.~\ref{fig:argmax}, where there is a transition in the arg max, at (b).  By contrast, taking the mean of the distribution ignores the bimodal nature of the distributions and shows no transition.  This argument also applies to observables, such as work, that are functions of filter estimates.

To know when keeping a history of observations pays off, we need to determine under what conditions the two state estimates will differ.  We quantify how similar two sequences of state estimates by defining a \textit{discord} order parameter,
\begin{equation}
	D = 1 - \frac{1}{N} \sum_{t=1}^N d( \hat{x}_t^{\rm o}, \hat{x}_t^{\rm f} ) \,,
\label{eq:discord_NN}
\end{equation}
where the function $d$ depends on the naive and filter state estimates:
\begin{equation}
	d( \hat{x}_t^{\rm o}, \hat{x}_t^{\rm f} ) =
 	 \begin{dcases}
		\; \; \; 1,	& \hat{x}_t^{\rm o} = \hat{x}_t^{\rm f} \\
		-1,		& \hat{x}_t^{\rm o} \neq \hat{x}_t^{\rm f}
  	\end{dcases} \,.
\end{equation}
The discord parameter is zero when the state estimates agree at all times.
In such a case, there is no value in keeping a memory of observations: the extra information contained in the past observations has not changed the best estimate from that calculated using only the present observation. Similarly, when $D=2$ the state estimates \textit{disagree} at all times, the state estimates are perfectly anti-correlated. At intermediate values of $D$, keeping a memory can be beneficial.   An HMM with a non-zero discord is illustrated in Fig.~\ref{fig:timeseries}:  an arrow indicates a point where the state estimate differs from the estimate based on the current observation.
    
We are interested in the transition from zero to non-zero discord, where the state estimates start to differ, and where keeping a history of observations starts to pay off.  The lowest observation probability that leads to a non-zero discord is the \textit{critical observation probability}.  We have just seen that after a long sequence of identical observations the probability to be in some state $x_t$ reaches a maximum value. Thus, the first place where state estimates will differ is when a single discordant observation after a long string of identical observations does \textit{not} change our belief of the state of the system (i.e., where the filter estimate no longer follows the naive estimate exactly).  Mathematically, the threshold where the discord goes from zero to being non-zero for an $n \times n$ HMM is given by
\begin{equation}
  \begin{aligned}
  \label{eq:bcdef}
		\lim_{t \to \infty} P(x_{t+1} &= i|y_{t+1} = j, y^t = i^t)\\
    &= \lim_{t \to \infty} P(x_{t+1} = j|y_{t+1} = j, y^t = i^t) \,.
  \end{aligned}
\end{equation}
for all states $i,j \in \{ 1, 2, \dots, n \}, \text{ and } j \neq i$.  From Ref.~\cite{bechhoefer2015hidden}, the transition threshold for a symmetric $2 \times 2$ HMM with transition probability $a$ and error rate $b$ is 
\begin{equation}
	b_{\rm c} = \tfrac{1}{2} \left( 1-\sqrt{1-4a} \right) \quad  \left( a \le \tfrac{1}{4} \right) \,,
\label{eq:2x2threshold}
\end{equation}
and $b_{\rm c}= 1/2$ for larger $a$ values. In Sec.~\ref{sec:22HMMs}, we generalize this result by dropping the symmetry requirement.
As found in \cite{bechhoefer2015hidden}, the transitions are sometimes discontinuous and sometimes just have a discontinuity in their derivative. As far as we know, the distinction has not been explored.

So far, we have only considered the extreme cases of no memory and infinitely long memory.
What about a finite memory?  In Fig.~\ref{fig:timeseries}, we see that the filter reacts to new observations with a characteristic timescale.  Indeed, since the filter dynamics for a system with $n$ internal states is itself a dynamical system with $n-1$ states (minus one because of probability normalization), we expect filters to have $n-1$ time scales.  This statement holds no matter how big or small the memory of the filter.

As a numerical exploration confirms, there is geometric (exponential) relaxation with time scales that are easy to evaluate numerically, if difficult algebraically.  Thus, an ``infinite" filter memory need only be somewhat longer than the slowest time scale, and ``no memory" need only be faster than the fastest time.  A brief study of filters with intermediate time scales suggests that their behavior typically interpolates between the two limits.

\section{Symmetry breaking in two-state, two-symbol HMMs} 
\label{sec:22HMMs}

In $2 \times 2$ HMMs where the symmetry in the transition and observation probabilities is broken, the probability matrices each have two independent parameters.   We parametrize the transition-matrix probabilities as
\begin{equation}
	\label{eq:matrixasymm}
	\mathbf{A} = 
	\begin{pmatrix}
		1-\bar{a} + \frac{1}{2}\Delta a & \bar{a} + \frac{1}{2} \Delta a  \\[10pt]
		\bar{a} - \frac{1}{2}\Delta a & 1-\bar{a} - \frac{1}{2} \Delta a
	\end{pmatrix} \,,
\end{equation}
which depends on the mean transition probability $\bar{a} = \frac{1}{2}(A_{21} + A_{12})$ and the difference in transition probabilities $\Delta a = A_{21} - A_{12}$.  When the difference between the transition probabilities is zero $(\Delta a = 0)$, the transition matrix is symmetric.  The observation matrix is parameterized similarly, with $\bar{a} \to \bar{b}$ and $\Delta a \to \Delta b$.  The matrix $\mathbf{B}$ depends on the mean observation probability $\bar{b} = \frac{1}{2}(B_{21} + B_{12})$ and the difference in observation probabilities $\Delta b = B_{21} - B_{12}$.  All matrix elements must be in the range $[0,1]$ to ensure proper normalization.  We restrict the off-diagonal elements (probability to transition to a different state or probability to make a wrong observation) to be $< 0.5$, to preclude anticorrelations.  We use the set $\{1,-1\}$ to label both states and the corresponding symbols for $2 \times 2$ HMMs.
    
\begin{figure}[t]
	\centering
	\includegraphics[width=0.47\textwidth]{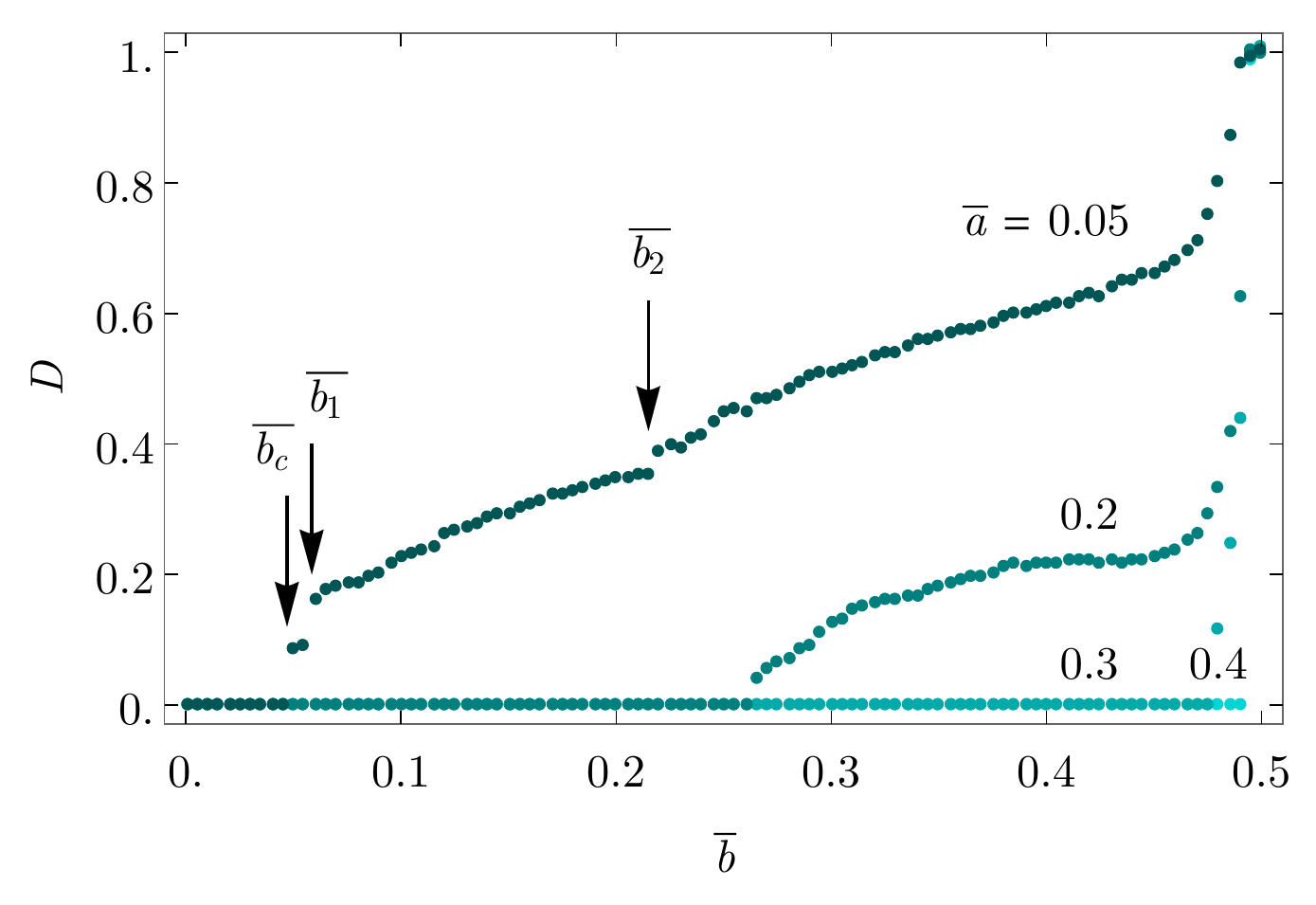}
	\caption{(Color online) Discord parameter as a function of the average observation probability for a $2 \times 2$ HMM with symmetric observations ($\Delta b = 0$) and slightly asymmetric transitions, $\Delta a = 0.01$. Arrows indicate the primary and two higher-order transitions.}
\label{fig:discord_asymmA}
\end{figure}

The discord parameter is calculated by generating a realization of an HMM using the transition and observation matrices, following Eq.~\eqref{eq:discord_NN} and averaging over the entire chain.  A plot of the discord for a $2 \times 2$ HMM with asymmetric transition probabilities is shown in Fig.~\ref{fig:discord_asymmA}. All points shown are averaged over $30,000$ time steps.

\begin{figure}[t]
	\centering
	\includegraphics[width=0.47\textwidth]{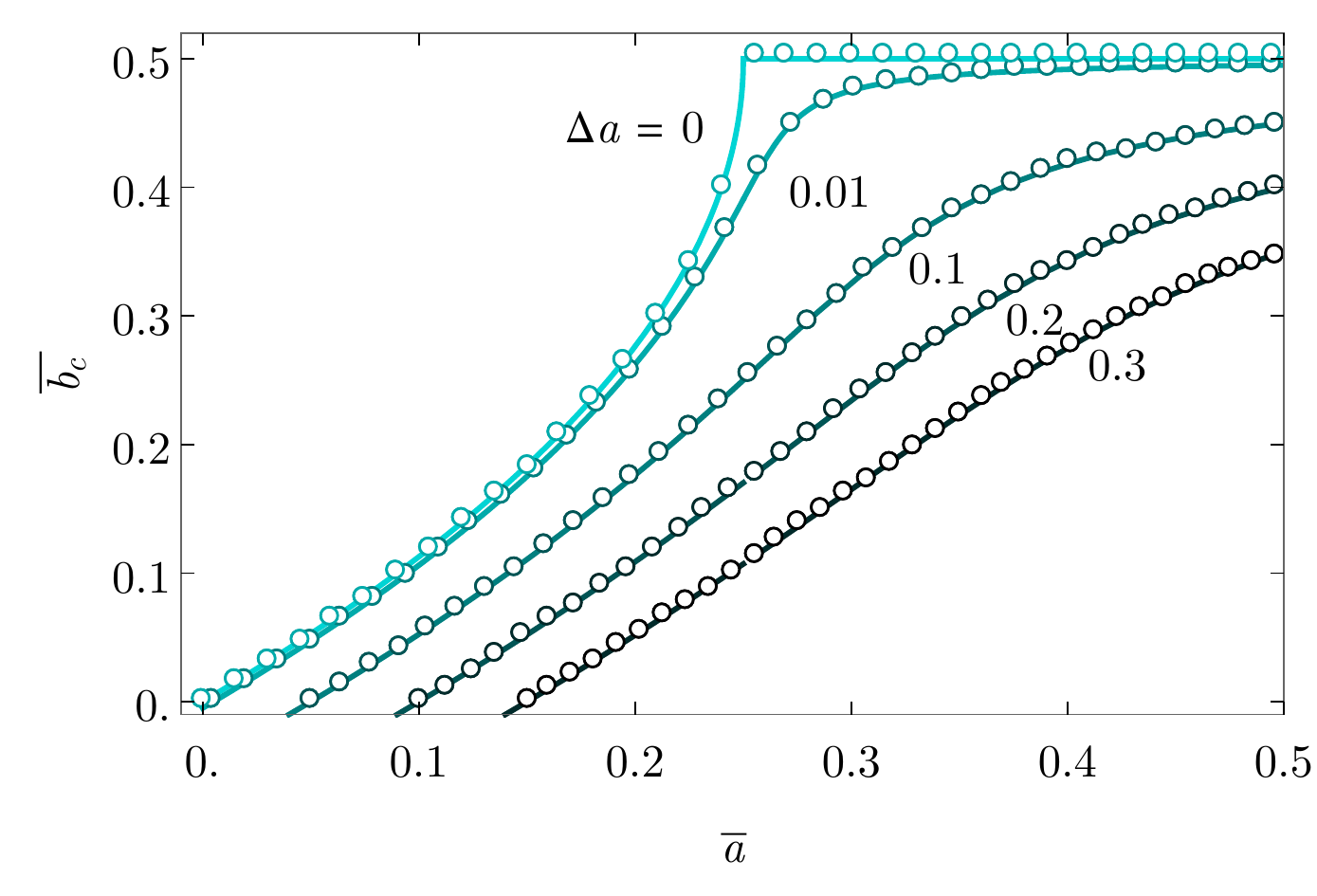}
	\caption{(Color online) Mean critical observation probability of HMMs with asymmetric transition matrices and symmetric observation matrices ($\Delta b = 0$) a function of $\bar{a}$. Simulated results are shown as circles; solid lines show the analytical solutions.}
\label{fig:bc_asymmA}
\end{figure}

The expression defining the critical observation probability in Eq.~\eqref{eq:bcdef} simplifies greatly for $2 \times 2$ HMMs:
\begin{equation}
	\lim_{t \to \infty} P(x_{t+1} = 1|y_{t+1} = -1, y^t = 1^t) = 0.5 \,.
\label{eq:bc2Ddef}
\end{equation}
We write this in terms of the model's parameters and solve for the critical observation probability $\bar{b} = \bar{b}_{\rm c}$.  
This corresponds to the lowest points $\bar{b}$ in Fig.~\ref{fig:discord_asymmA} that are non-zero for a given $\bar{a}$.  
The complete analytical calculations for the critical observation probability can be found in App.~\ref{app:bc2x2}.  
Figure~\ref{fig:bc_asymmA} shows the analytical and simulated critical observation probabilities as a function of the mean transition probability $\bar{a}$, for a system with symmetric observation probabilities and several different transition asymmetries.  
The curve labeled $\Delta a = 0.01$ corresponds to the transitions in Fig.~\ref{fig:discord_asymmA}.  
The solutions agree with simulations, which are shown as circles in the same diagram.  
The discord becomes non-zero at lower mean transition probabilities for larger asymmetries.
The phase transitions differ in location from those of the symmetric $2 \times 2$ HMMs, but they still exist.  
We find similar results in systems with symmetric transition matrices and asymmetric observation matrices, and in systems with both asymmetric transition and observation matrices \cite{lathouwers16}.

Another approach to understanding these results is offered in App.~\ref{app:ising}, where we show that we can map $2 \times 2$ HMMs onto one-dimensional Ising models with disordered fields and zero-temperature phase transitions.

In Fig.~\ref{fig:discord_asymmA}, we observe some additional jumps and kinks at error rates $\left( \bar{b} > \bar{b}_{\rm c} \right)$ that can be interpreted as ``higher-order transitions'' in the discord.  We have labeled two of such transitions by $\bar{b}_1$ and $\bar{b}_2$ in Fig.~\ref{fig:discord_asymmA}.  The first of these is due to the asymmetry of this HMM.  The threshold $\bar{b}_{\rm c}$ results from the observation sequence $y^t = \{1, \ldots , 1, -1 \}$, whereas the slightly higher $\bar{b}_1$ results from the sequence $y^t = \{-1, \ldots , -1, 1 \}$, which gives a condition that is different when $\Delta a \neq 0$.  The second of these transitions, $\bar{b}_2$, marks the threshold where two discordant observations are needed to change the filter state estimate.  That is, the observation sequence is $y^t = \{1, \dots, 1, -1, -1 \}$.  For still higher values of $\bar{b}$, there will be transitions where one needs more than two sequential discordant observations to alter the filter value.  Further transitions can occur for finite arbitrary sequences, too.  Higher-order transitions, however, are increasingly weak and harder to detect numerically.

\section{More states and symbols} 
\label{sec:nnHMMs}
    
We have seen that phase transitions in the discord order parameter occur in both symmetric and asymmetric time-homogeneous $2 \times 2$ HMMs.  In this section, we will study systems with more states and more symbols.  To keep the number of parameters manageable, we will consider symmetric HMMs, and we will consider only two classes of states:  an observation is either correct and the symbol is the ``same'' as the underlying state, or an observation is incorrect and the system emits an ``other'' symbol.  We consider a straightforward generalization of symmetric $2 \times 2$ HMMs to symmetric $n \times n$ HMM, and a model that describes a particle diffusing on a lattice with constant background noise.  We investigate whether the transitions that we have encountered so far exist in these systems, too.

\subsection{Symmetric \texorpdfstring{$n \times n$}{n} HMMs}
    
Let us now label the states and symbols $1, 2, \dots, n$.  We first consider a system with transition matrix
\begin{align}
	\mathbf{A} =
	\begin{pmatrix}
		1-a 	& \frac{a}{n-1} 	& \dots	 & \frac{a}{n-1}	\\[5pt]
		\frac{a}{n-1} & 1-a  & \ddots & \vdots 	\\[5pt]
		\vdots & \ddots	& \ddots & \frac{a}{n-1}	\\[5pt]
		\frac{a}{n-1}	& \dots	& \frac{a}{n-1}	& 1-a
	\end{pmatrix} \,,
\label{eq:NNmatrix}
\end{align}
and an observation matrix $\mathbf{B}$, which has the same form except that $a \to b$.  This system depends on only two parameters for a given number of states $n$: the transition probability, $a$, and the observation error probability, $b$.   This transition matrix describes a system that has a probability $1-a$ to stay in the same state and equal probabilities to transition to any other state, $\frac{a}{n-1}$.  The observation matrix describes a measurement with uniform background noise; there is a certain probability of observing the correct symbol $1-b$ and equal probabilities of observing any other symbol, $\frac{b}{n-1}$.
    
Just as before, we calculate the discord parameter for these systems and study the transition to non-zero discord by finding the critical observation probability. The problem simplifies from the case discussed above, where $\mathbf{A}$ is a general transition matrix. In App.~\ref{app:bc}, we write it out explicitly, and find two solutions:

\begin{subequations}
\label{eq:threshold_nstates}
	\begin{align}
	\label{eq:threshold_nstates1}
	b_{\rm c}^{(1)} 	&= \frac{1}{2(n-1)} \Big( (n-1) + (n-2) a - \Big.  \\ 
    				& \Big. \sqrt{(n-2)^2 a^2 - 2 n (n-1) a + (n-1)^2} \Big) \,, \notag \\
	b_{\rm c}^{(2)} 	&= \frac{n-1}{n} \,.
	\label{eq:threshold_nstates2}
	\end{align}
\end{subequations}
\begin{figure}[t]
	\centering
	\includegraphics[width=0.47\textwidth, keepaspectratio]{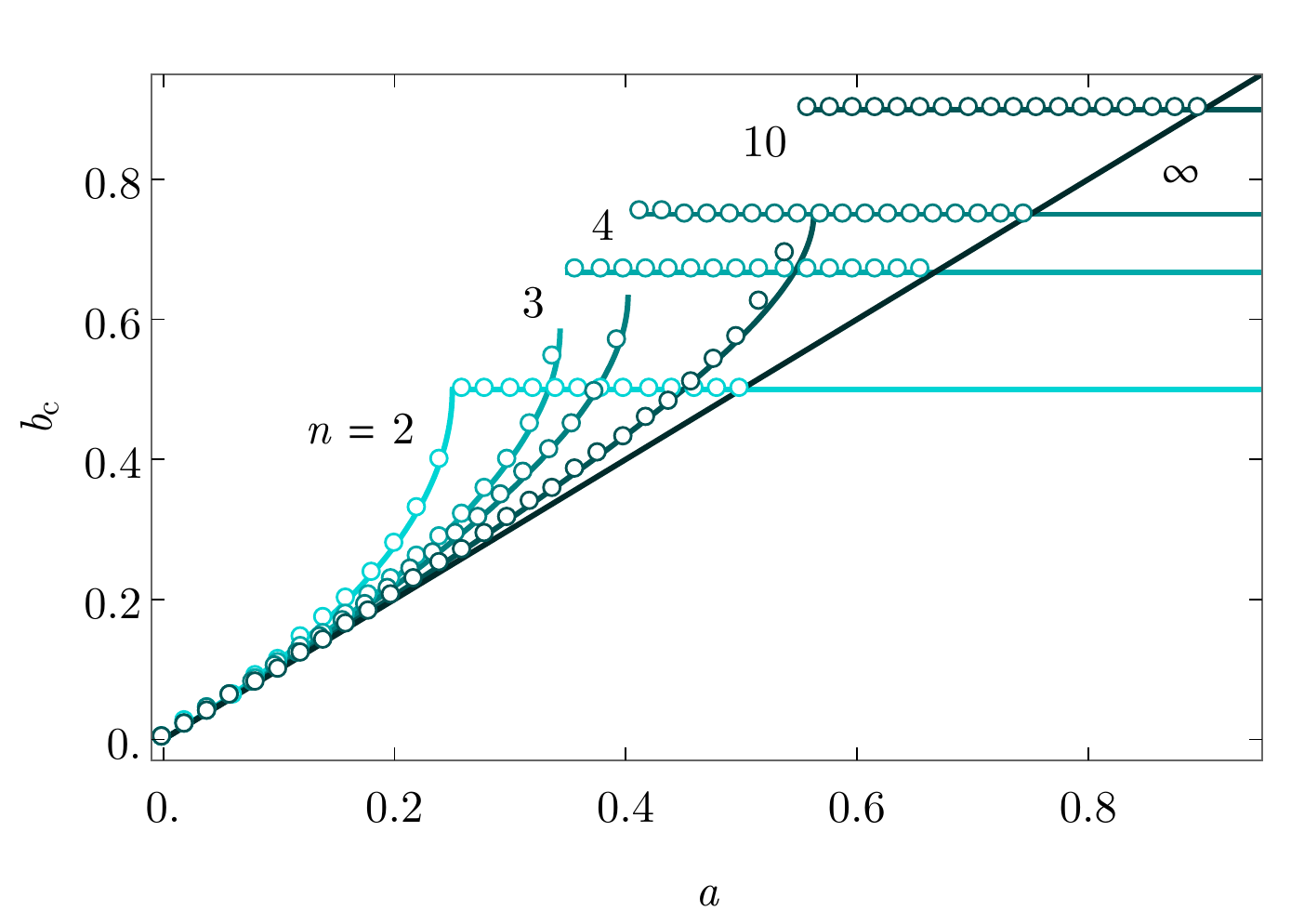}
	\caption{(Color online) Phase diagram of symmetric $n \times n$ HMMs for $n = 2, 3, 4, 10$, and $n \to \infty$.  The lines are analytical solutions; the circles are the results of simulations.}
\label{fig:PDnDsymm}
\end{figure}
For $n=2$, Eq.~\eqref{eq:threshold_nstates1} reduces to Eq.~\eqref{eq:2x2threshold}.

The threshold values of $b$ are plotted in Fig.~\ref{fig:PDnDsymm} for various numbers of states and symbols $n$. 
The branches of the solutions that are increasing with increasing $a$ are given by $b_{\rm c}^{(1)}$, and the constant branches are given by $b_{\rm c}^{(2)}$.
The analytical and simulated values agree quite well, especially for smaller $n$.  The $n = 10$ curve deviates from the simulations slightly at higher $a$.  The area under the curves indicates the parameter regime where $D=0$, where the state estimates with and without memory agree.  Above the critical error probability, the two state estimates differ.  There are no discontinuities as $b_{\rm c} \to \frac{n-1}{n}$; the curves are simply very steep.

\subsection{Diffusing particle}
      
We now consider an HMM that describes a particle diffusing on a lattice with constant background noise and periodic boundary conditions.
The symmetric $n \times n$ transition matrix is given by
\begin{equation}
	\mathbf{A} =
	\begin{pmatrix}
		1-a 	& \frac{a}{2} 	& 0	& \dots	& \frac{a}{2}	\\[5pt]
	  	\frac{a}{2} 	& 1-a	& \frac{a}{2}	& \ddots	& 0	\\[5pt]
		0	& \frac{a}{2}	& 1-a	& \ddots	& \vdots	\\[5pt]
		\vdots	& \ddots	& \ddots	& \ddots	& \frac{a}{2}	\\[5pt]
		\frac{a}{2}	& 0	& \dots	& \frac{a}{2}	& 1-a	
	\end{pmatrix} \,.
\end{equation}
The observation matrix is the same as the one in the previous section.
Physically, the particle stays in the same place with probability $1-a$ or it diffuses one site to the left or right with probability $a/2$.

\begin{figure}[t]
	\centering
	\includegraphics[width=.48\textwidth]{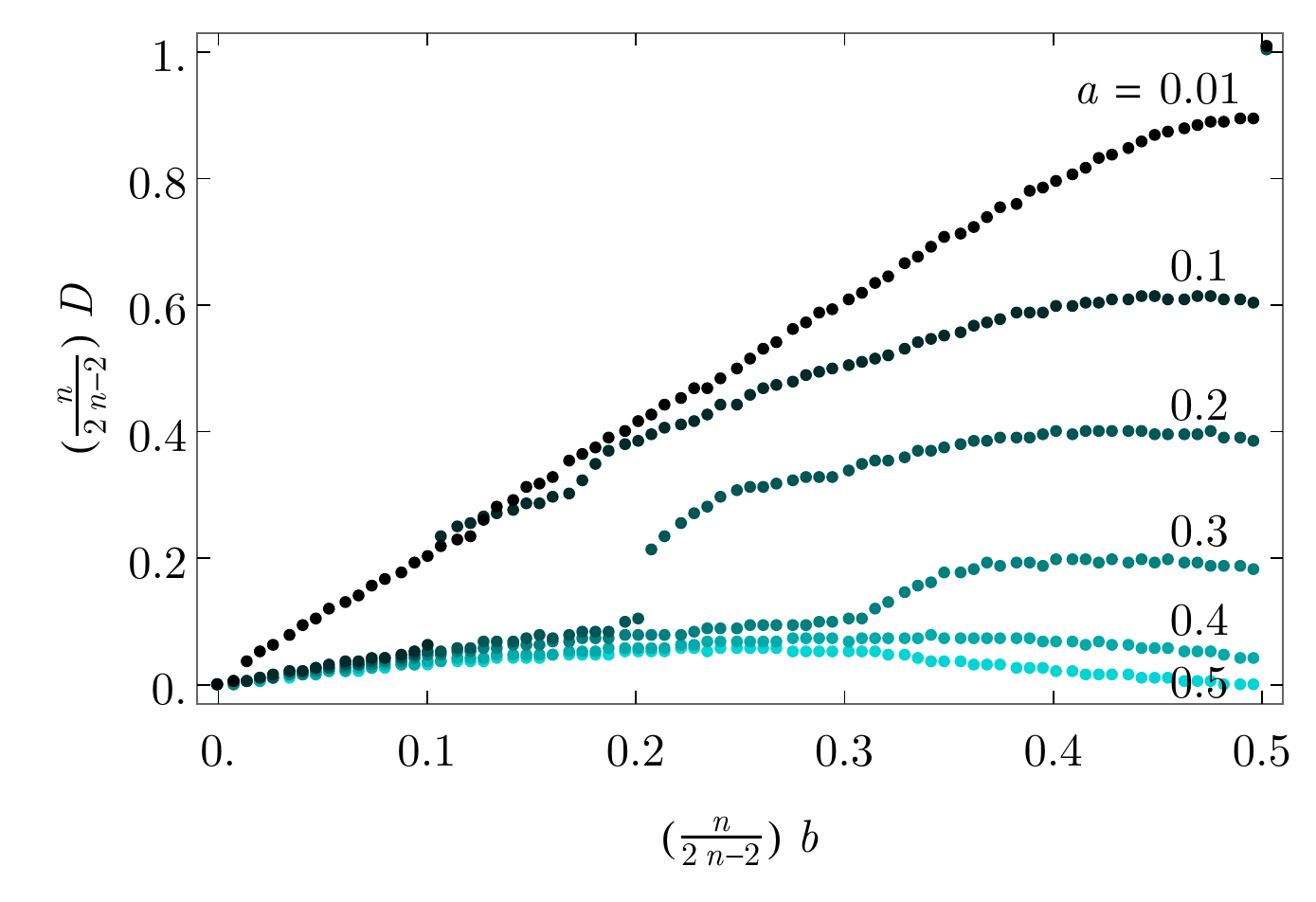}
	\caption{(Color online) Scaled discord parameter as a function of the scaled observation probability for a diffusing particle on a one-dimensional lattice with four sites ($n=4$).}
\label{fig:Discord_DP}
\end{figure}

The discord parameter as a function of the observation probability is plotted in Fig.~\ref{fig:Discord_DP}.
For visualization purposes, both the discord and the observation probability are scaled by a factor of $n/(2n-2)$, where $n$ is the number of lattice sites.
The scaling is such that $(n/(2n-2))D = 1$ at $(n/(2n-2))b = 0.5$ for any integer $n > 1$.
The transition to non-zero discord is smooth in this case; however, at higher $b$ and $D$, a non-analytic jump is seen.

\section{More symbols than states} 
\label{sec:2nHMMs}

Finally, we consider an HMM with more symbols than states.  In particular, consider an HMM with only two states, 1 and $-1$, and an even number of symbols $n$.  We will also consider the $n \to \infty$ limit.  The transitions and errors are once again taken to be symmetric, with $\mathbf{A} = \bigl( \begin{smallmatrix} 1-a & a \\ a & 1-a \end{smallmatrix} \bigr)$, but the observation errors are now determined by a Gaussian distribution around the states.  In particular, the elements of the observation matrix are determined by integrals over the Gaussian distribution of the desired state.  For state 1, the observation probability for symbol $i$ is
\begin{align}
	b_{i1} &= P(y_t = i|x_t = 1) \notag \\[5pt]
	&= \frac{1}{\sqrt{2 \pi} \sigma} \, \int_{\ell_{i-1}}^{\ell_i} \mathrm{d}x \, 
		\exp \left( \frac{-(x - 1)^2}{2 \sigma^2} \right) \notag \\[5pt]
	&= \frac{1}{2} \left[ \erf \left( \frac{\ell_i - 1} {\sqrt{2} \sigma} \right) - 
		\erf \left( \frac{\ell_{i-1} - 1}{\sqrt{2} \sigma} \right) \right] \,. 
\end{align}
The boundaries $\ell_i$ of the integral are determined such that the probability of observing each symbol is equal when considering the sum of Gaussian distributions around each state.
\begin{align}
	\frac{i}{n} &= \frac{1}{\sqrt{8 \pi} \sigma} \, \int_{- \infty}^{\ell_i} \mathrm{d}x \, 
		\exp {\left( \frac{-(x - 1)^2}{2 \sigma^2} \right)} 
		+ \exp {\left( \frac{-(x + 1)^2}{2 \sigma^2} \right)} \notag \\[5pt]
	&= \frac{1}{4} \left[ 2 + \erf \left( \frac{\ell_i - 1}{\sqrt{2} \sigma} \right) 
		+ \erf \left( \frac{\ell_i + 1}{\sqrt{2} \sigma} \right) \right] \,.
\end{align}
The symmetry of the problem reduces the number of equations we need to solve: We know that $\ell_0 = -\infty$, $\ell_n = \infty$, $\ell_{n/2} = 0$, and $\ell_{(n/2)-j} = - \ell_{(n/2)+j}$ for integers $j$ between 1 and $n/2-1$ (all for even $n$).  Also, symmetry dictates that the probability of observing a symbol $i$ given the state is $-1$, $b_{i(-1)}$, equals $b_{(n-i+1)1}$.  

For $n \to \infty$ (infinitely many symbols), we use the probability density function directly rather than integrating over an interval.
  
\begin{figure}[t]
	\centering
	\includegraphics[width=.4\textwidth]{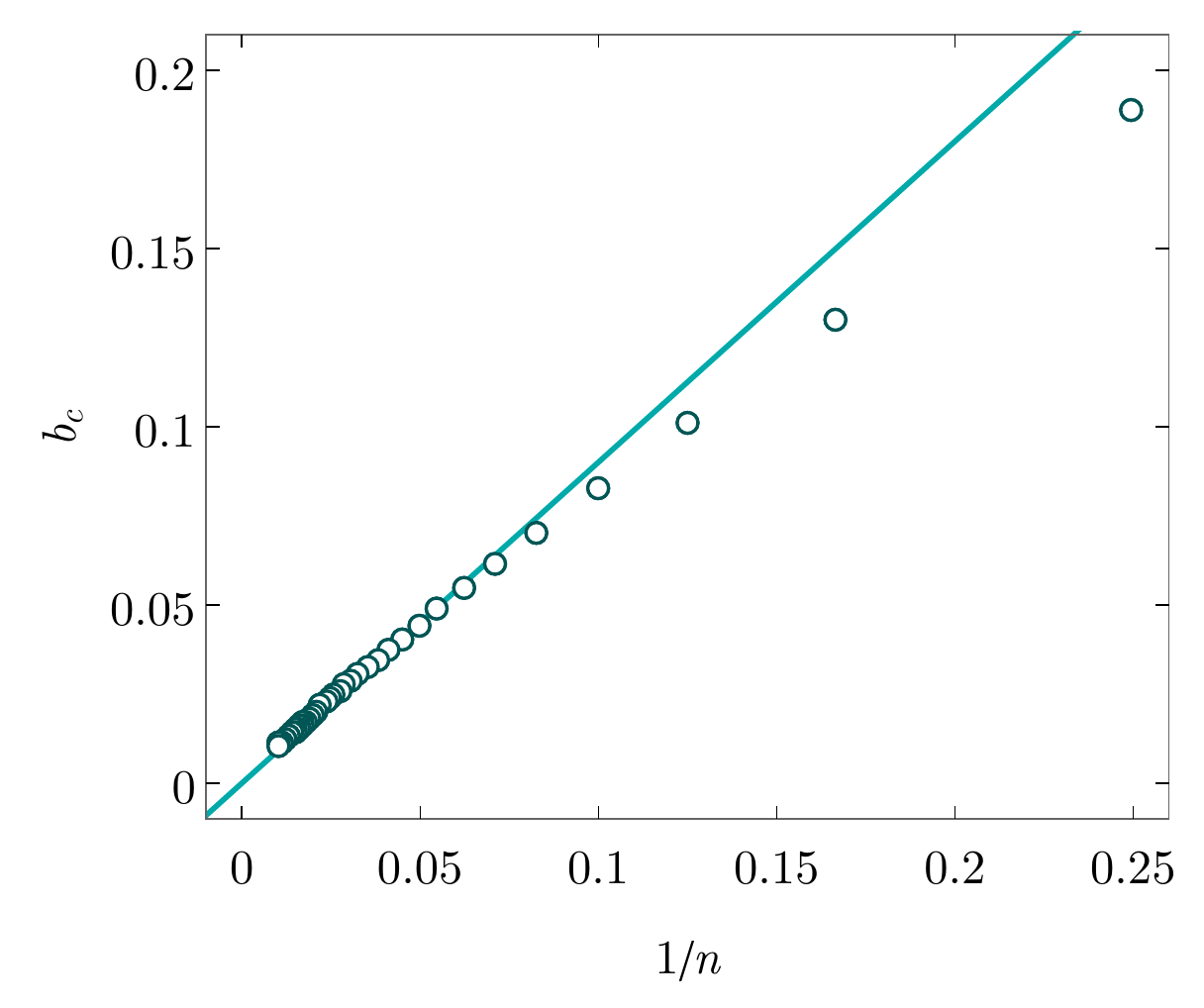}
	\caption{(Color online) Critical observation probability $b_{\rm c}$ of symmetric $2 \times n$ HMMs, for a fixed transition probability $a = 0.30$, as a function of $1/n$. The straight line emphasizes the asymptotic behavior for large $n$.}
\label{fig:PD_2n}
\end{figure}

In systems with a finite number of symbols, we observe non-analytic behavior as the discord becomes non-zero.  These phase transitions move to lower observation probabilities for a larger number of symbols.  In systems with an infinite number of symbols, the discontinuities are not present.  To confirm this observation, we study the critical error probability as a function of the number of symbols; see Fig.~\ref{fig:PD_2n}.
The critical error probability is shown for a transition probability $a = 0.30$ as a function of $1/n$. The behavior is similar for other transition probabilities and suggests that the critical observation probability goes to zero asymptotically.  The (discontinuous) transitions disappear only in the limit of infinitely many symbols.
  
In the $n \times n$ symmetric HMMs, we saw similar behavior: The critical error probability decreases as a function of the number of symbols (and states).  However, if we look at the limit of $n \to \infty$ of the critical error probability of these HMMs (Fig.~\ref{fig:PDnDsymm}), the critical error probability does not go to zero.

\section{Conclusion} 
\label{sec:conclusion}

In this paper, we have investigated when keeping a memory of observations pays off in hidden Markov models.  We used HMMs to look at a relatively simple system with discrete states and noisy observations.  We inferred the underlying state of the HMM from the observations in two ways: through naive observations, and a state estimate found through Bayesian filtering (and decision making).  
We then compared the state estimates by calculating the discord, $D$, between the two.  We were particularly interested in investigating a phase transition at the point where $D$ becomes non-zero.  Such transitions have been observed in symmetric $2 \times 2$ HMMs; here, we have seen that such behavior applies to more general models.
  
We looked at asymmetric $2 \times 2$ HMMs, some symmetric $n \times n$ HMMs, and symmetric $2 \times n$ HMMs.  The general features of the discord stayed the same in all these systems: it starts at $D=0$ for $b=0$; it becomes non-zero at some critical error probability; and it increases for increasing error probability.  In all these systems, we found a non-analytic behavior in the discord as a function of observation error probability (phase transition), except in the $2 \times n$ case in the limit of infinitely many symbols, $n \to \infty$.

Throughout this paper, we have defined the usefulness of memory in a rather narrow way:  we ask when inferences using a memory are different or better than those that do not.  But memory can have many more uses.  In thermodynamics, Maxwell's demon and Szilard's engine showed that information that is acquired can be converted to work \cite{parrondo15}.  Bauer et al. analyzed a periodic, two-state Maxwell demon with noisy state measurements and showed that there are transitions very much analogous to the ones considered here between phases where measurements are judged reliable, or not \cite{bauer14}.  When reliable, there is no advantage to keeping a memory.

In biology, one can consider cells in noisy environments and ask whether keeping a memory of observations of this environment is worthwhile.  For example, Sivak and Thomson showed, in a simple model, that for very low and very high ratios of signal to noise in the environment, memoryless algorithms lead to optimal regulatory strategies \cite{sivak14}.  However, for intermediate levels of noise, strategies that retain a memory perform better.  In contrast to the situations considered in this paper, they found no evidence of any phase transition.  There were smooth crossovers between regimes. In another setting, Rivoire and Leibler have explored how information retention by \textit{populations} of organisms can improve the ability of the population to adapt to fluctuating environment \cite{rivoire11}.  Again, in this setting, no phase transitions were encountered.
Also, Hartich et al. showed that the performance of a sensor, characterized by its ``sensory capacity,'' increases with the addition of a memory but report no  phase transition \cite{hartich2016sensory}. 

Thus, in this paper, we have explored a class of models where phase transitions occur generically as a function of signal-to-noise ratios.  Yet, in many other applications, such transitions are not observed.  Clearly, a better understanding is needed to clarify which settings will show phase transitions and which ones continuous crossovers between different regimes.
 
\appendix
\begin{widetext}

\section{Critical error probability of asymmetric \texorpdfstring{$2 \times 2$}{2} HMMs}
\label{app:bc2x2}

In Eq.~\eqref{eq:bc2Ddef}, we defined the critical error probability $\bar{b}_{\rm c}$ as the lowest $\bar{b}$ for given $\bar{a}, \Delta a,$ and $\Delta b$ that results in a non-zero discord parameter.  In this appendix, we calculate this threshold analytically.

We start by writing out the left-hand-side of Eq.~\eqref{eq:bc2Ddef} explicitly:
\begin{equation}
\label{eq:bcasymm}
	\lim_{t \to \infty} P(x_{t+1} = 1|y_{t+1} = -1, y^t = 1^t) = \lim_{t \to \infty}  \frac{ P(y_{t+1} 
		= -1|x_{t+1} = 1) \sum_{x_t} P(x_{t+1} = 1|x_t) P(x_t|y^t = 1^t)}
		{ \sum_{x_{t+1}} P(y_{t+1} = -1|x_{t+1})  \sum_{x_t} P(x_{t+1}|x_t)  P(x_t|y^t = 1^t)  }  \,.
\end{equation}
We recognize several terms as part of the transition and observation matrices. The term $P(x_t|y^t=1^t)$ relates to the maximum confidence level.  We need to find the maximum confidence level in terms of the observation and transition probabilities.
\begin{align} 
	p_1^* &= \lim_{t \to \infty} P(x_t = 1|y^t = 1^t) \notag \\
				&= \lim_{t \to \infty}  \frac{ P(y_t = 1|x_t = 1) \sum_{x_{t-1}} P(x_t = 1|x_{t-1}) P(x_{t-1}|y^{t-1} = 1^{t-1}) } { \sum_{x_{t}} P(y_{t} = 1|x_{t}) \sum_{x_{t-1}} P(x_{t}|x_{t-1}) P(x_{t-1}|y^{t-1} = 1^{t-1}) } \, .
\end{align}
We use normalization to write $\lim_{t \to \infty} P(x_{t-1} = -1|y^{t-1} = 1^{t-1}) = 1 - p^*_1$, and we have an expression only in terms of the maximum confidence level, transition probability and the observation probability.
We then solve for $p^*_1$:
\begin{align}
 p_1^* &= \frac{1}{4(2\bar{a} - 1)(2\bar{b} - 1)} \left( 2 + \bar{a}(8\bar{b} - 2(3 + \Delta b)) + 2\bar{b}(\Delta a - 2) - \Delta a	 + X \right), \text{ with} \\ 
 X &= \sqrt{4\bar{a}^2 (1 + \Delta b)^2 - 4\bar{a} (2\bar{b} - 1)(4\bar{b} - 2 - \Delta a (1 + \Delta b)) + (2\bar{b} - 1) \left( 2\bar{b}(4 + \Delta a^2) - \Delta a (4 + \Delta a + 4 \Delta b) - 4 \right) } \, .
\end{align}
Now we plug this expression, together with the transition and observation probabilities (Eq.~\eqref{eq:matrixasymm}) into Eq.~\eqref{eq:bcasymm}: 
\begin{equation}
	\frac{(\Delta b - 2\bar{b}) \left( (2\bar{b} - 1)(2 + \Delta a) + 2\bar{a} (1 + \Delta b)  - X \right)}{2(2\bar{b} -1) \left( \Delta a(1 - 2\bar{b}) + 2 (\Delta b - 1) -2\bar{a}(1 + \Delta b) + X \right)} = \frac{1}{2} \, .
\end{equation}
Lastly, we solve for $\bar{b} = \bar{b}_{\rm c}$. We find three solutions, of which only two lie in our region of interest, $0 \leq \bar{a},\bar{b} \leq 0.5$. Since the resulting expressions are complicated, we show the full solution only for the special case where $\Delta b = 0$:
\begin{align}
	\bar{b}_{\rm c}^{(1)} &= \frac{1}{6} \left( 3 - \Delta a + \frac{3 - 12 \bar{a} + \Delta a^2}{ Y } + Y \right) \, , \\
	\bar{b}_{\rm c}^{(2)} &= \frac{1}{384} \left( -64(\Delta a - 3) - \frac{32 \left( 1 + i \sqrt{3} \right) (3 - 12 \bar{a} + \Delta a^2)}{ Y } +  32 i \left( i + \sqrt{3} \right) Y \right) \, , \\
	Y &= \left( 18 (\bar{a}-1) \Delta a - \Delta a^3 + 3 \sqrt{3} \sqrt{(11-4 \bar{a} (\bar{a}+4)) \Delta a^2 + (4 \bar{a}-1)^3 + \Delta a^4} \right)^{1/3} \, .
\end{align}
These solutions are plotted in Fig.~\ref{fig:bc_asymmA}.  Note that the expression for $\bar{b}_{\rm c}^{(2)}$ is real for relevant branches.  That is, for some values of $\bar{a} \text{ and } \Delta a$ the expression is complex; however, all the branches we plot have a zero imaginary part. When we set $\Delta a = 0$, $b_c^{(1)}$ reduces to $\frac{1}{2} \left( 1 - \sqrt{1 - 4\bar{a}} \right)$ for $\bar{a} \leq 1/4$, and $\frac{1}{2}$ for $\bar{a} \geq 1/4$. These are the familiar solutions for symmetric $2 \times 2$ HMMs as found in \cite{bechhoefer2015hidden} and App.~\ref{app:bc}.

\section{Mapping to Ising models}
\label{app:ising}
    
One can map a symmetric $2 \times 2$ HMM onto a one-dimensional random-field Ising model \cite{zuk2005entropy, allahverdyan2009maximum, bechhoefer2015hidden}.  Here, we generalize this mapping so that it applies to a general (asymmetric) $2 \times 2$ HMM.  We start by defining a mapping from the transition and observation probabilities to the spin-spin coupling and the spin-field coupling constants,
\begin{align}
	P(x_{t+1}|x_t) &= \frac{\exp(J(x_t) x_{t+1} x_t)}{2 \cosh(J(x_t))} \,, 
	&J(x_t) &= 
	\begin{dcases}
	  J_+ = \frac{1}{2} \log{\left( \frac{1-\bar{a} + \Delta a/2 }{\bar{a} 
	  	- \Delta a/2} \right)},& \text{if } x_t = 1 \\
	  J_- = \frac{1}{2} \log{\left( \frac{1-\bar{a} - \Delta a/2 }{\bar{a} 
	  	+ \Delta a/2} \right)},& \text{if } x_t = -1  
	\end{dcases} \,, \notag \\
	P(y_t|x_t) &= \frac{\exp(h(x_t) y_t x_t)}{2 \cosh(h(x_t))} \,, 
	&h(x_t) &=
	\begin{dcases}
	  h_+ = \frac{1}{2} \log{\left(\frac{1-\bar{b} + \Delta b/2 }{\bar{b} 
	  	- \Delta b/2}\right)},& \text{if } x_t = 1\\
	  h_- = \frac{1}{2} \log{\left(\frac{1-\bar{b} - \Delta b/2 }{\bar{b} 
	  	+ \Delta b/2}\right)},& \text{if } x_t = -1 
	\end{dcases} \,. 
\label{eq:asymm_map}
\end{align}
We define the Hamiltonian $\mathcal{H} \equiv -\log(P(x^N, y^N))$, which, using the product rule of probability and the Markov property of the state dynamics, is
\begin{align}
	\mathcal{H} = &-\sum_{t=1}^{N} \log{\left(P(y_t|x_t)\right)} -\sum_{s=1}^{N-1} \log{\left(P(x_{s+1}|x_s)\right)} \notag \\
	= &-\sum_{t=1}^{N} \left[ h(x_t) y_t x_t - \log(2 \cosh(h(x_t))) \right] -\sum_{s=1}^{N-1} \left[ J(x_s) x_{s+1} x_s - \log(2 \cosh(J(x_s))) \right] \,.
\end{align}
Next, we rewrite the $h(x_t)$ and $J(x_t)$ in a convenient way:
\begin{equation}
  \begin{aligned}
		h(x_t) &=  \bar{h} + \Delta h x_t, \quad \text{with} \quad
		\bar{h} = \frac{1}{2}(h_+ + h_-) \quad \text{and} \quad \Delta h = \frac{1}{2}(h_+ - h_-) \,,
  \end{aligned}
\end{equation}
and the same for $J(x_t)$ with $h \to J$.  When $\Delta a$ is zero, we have $\Delta J = 0$ and $\bar{J} = J$, where $J$ is the coupling constant found in the case of symmetric $2 \times 2$ HMMs \cite{bechhoefer2015hidden}.  The same happens with the $h$-terms when $\Delta b = 0$.  The terms consisting of a logarithm with a hyperbolic cosine can also rewritten by taking the mean value of the possible terms and a deviation from that mean value.  The constant terms can be neglected since they lead only to a shift in the energy.  Similarly, terms that depend only on a single factor $y_t$ can also be neglected.  Higher-order terms that depend on a product of these factors still contribute. 

The full Hamiltonian is now given by
\begin{align}
	\mathcal{H} = &-\sum_{t=1}^{N} \left[ \bar{h} y_t x_t -\frac{1}{2}x_t\log{\left( \frac{\cosh(\bar{h}
		+\Delta h)}{\cosh(\bar{h}-\Delta h)} \right)} 	\right]
	 -\sum_{s=1}^{N-1} \left[ \bar{J} x_{s+1} x_s + \Delta J x_{s+1} 
		-\frac{1}{2}x_s\log{\left( \frac{\cosh(\bar{J}+\Delta J)}{\cosh(\bar{J}-\Delta J)} 
		\right)} \right] \,.
\end{align}
For large $N$, we can neglect boundary terms.  Then rearranging the Hamiltonian so that one term represents the nearest-neighbor interactions and the others the local external fields, we find,
\begin{align} 
	\mathcal{H} = & -\sum_t \bar{J} x_{t+1} x_t - \left[ \bar{h} y_t
		-\frac{1}{2} \log{\left( \frac{\cosh(\bar{h} + \Delta h)}{\cosh(\bar{h}-\Delta h)} \right)} 
		-\frac{1}{2} \log{\left( \frac{\cosh(\bar{J} + \Delta J)}
			 {\cosh(\bar{J}-\Delta J)} \right)} + \Delta J \right] x_t 		\notag \\
	  = & -\sum_t \bar{J} x_{t+1} x_t - \bar{h} y_t x_t 
	  	+ C(\bar{J}, \Delta J, \bar{h}, \Delta h) x_t \,.
\label{eq:Hising}
\end{align}
The external field consists of a fluctuating term that depends on $y_t$ and a constant term that depends transition and observation parameters.

\end{widetext}

From Eq.~\eqref{eq:Hising}, it is clear that this Hamiltonian remains the Hamiltonian of the familiar Ising model.  There is a constant spin-spin coupling term, the strength of which is determined by the transition probabilities $\bar{a}$ and $\Delta a$.  Then there is the fluctuating term of the local external fields.  The magnitude is constant and determined by the observation probabilities, but the direction is assigned randomly through $y_t$.  Finally, there is a constant term in the external fields that depends on both the transition and observation probabilities.

Above, we have seen that the filtering problem for a HMM can be mapped onto an Ising model.   How useful is such a mapping?  It does add intuitive language.  The observations $y_t$ play the role of a local spin at each site.  From Eq.~\eqref{eq:asymm_map}, we see that a lower error rate (small $\bar{b}$) corresponds to strong coupling between the local field and the local spin, which corresponds to state $x_t$.  When the noise is so strong that an observation says nothing about the underlying state ($\bar{b}=1/2$), then the coupling $h=0$.  

Likewise, deviations of $\bar{a}$ from 1/2 determine the spin-spin coupling constant $J$.

These results, however, were previously derived for symmetric $2 \times 2$ HMMs \cite{zuk2005entropy, allahverdyan2009maximum}.  Here, we add the insight that generalizing to asymmetric dynamics (matrix $\mathbf{A}$) or observation errors (matrix $\mathbf{B}$) leads to the same qualitative scenario.  The mapping remains a simple Ising model; only the coefficients are modified.  It would be interesting to know whether such mappings work for more order parameters, where the corresponding spin problem is presumably a Potts model \cite{wu82}.

Although the Ising mapping gives some qualitative insights, it has limitations.  In a closely related problem, estimating the \textit{full path} $x^t$ of states on the basis of observations $y^t$, the desired filter estimate corresponds to the ground state of the corresponding Ising model \cite{allahverdyan2009maximum}.  Here, by contrast, the filter estimate corresponds, in Ising language, to estimating the most likely value of the last (edge) spin of a 1d chain, without caring about the spin of any other site---a strange quantity!  
Thus, in mapping the filter state estimation problem to an Ising chain, we transform a familiar question concerning a strange system to asking a strange question of a familiar system.  Whether such a swap leads to analytical progress beyond its value in forming a qualitative picture is not at present clear.

\section{Calculation of critical observation probability for \texorpdfstring{$n \times n$}{n} HMMs} 
\label{app:bc}

Here, we compute the critical observation probability for symmetric $n \times n$ HMMs analytically.
We need only consider one state/ symbol $i$ and one $j$, thanks to the symmetry of the problem.
For example, use $i = 1$ and $j = 2$ in Eq.~\eqref{eq:bcdef}, which leads to
\begin{equation}
	\begin{aligned} 
		\lim_{t \to \infty} P(x_{t+1} &= 1|y_{t+1} = 2, y^t = 1^t) \\
		&= \lim_{t \to \infty}  P(x_{t+1} = 2|y_{t+1} = 2, y^t = 1^t) \,.
	\end{aligned}
\label{eq:bcdefnn}
\end{equation}

Similar to the calculation in App.~\ref{app:bc2x2}, we start with the calculation of the maximum confidence level, $p^*$.  Since all states of a symmetric $n \times n$ HMM are equivalent, the maximum confidence levels are all the same.  We calculate the maximum confidence for an arbitrary state $i$,
\begin{align}
	p^{*} = \lim_{t \to \infty} P(&x_t = i|y^t = i^t) \notag \\ 
		= \lim_{t \to \infty}  \frac{1}{Z_t} &P(y_t = i|x_t = i) \notag \\
		&\sum_{x_{t-1}} P(x_t = i|x_{t-1}) P(x_{t-1}|y^{t-1} = i^{t-1}) \,. 
\label{eq:def_maxconfnD}
\end{align}
The first two terms in the numerator are known from the transition and observation matrix of the HMM.  The last term is $p^{*}$ if $x_{t-1} = i$.   For $x_{t-1} \neq i$, we can calculate it by demanding a normalized probability,
\begin{align}
	\lim_{t \to \infty} \sum_{x_{t-1}} P(x_{t-1}|y^{t-1} = i^{t-1}) &= 1 \notag \\
	p^{*} + (n-1) \lim_{t \to \infty} P(x_{t-1} = j|y^{t-1} = i^{t-1}) &= 1 \notag \\[5pt]
	\lim_{t \to \infty} P(x_{t-1} = j|y^{t-1}=i^{t-1}) &= \frac{1 - p^*}{n-1} \,. 
\label{eq:not_pstar}
\end{align}
Plugging all of the terms into Eq.~\eqref{eq:def_maxconfnD} and solving for $p^*$ in terms of the model parameters leaves us with two solutions.  Restricting interest to the solutions that take on positive values for $0 \leq a, b \leq 1$ and integer $n>1$, we find,
\begin{widetext}

\begin{equation}
	\begin{aligned}
 		 p^{*} = &\frac{1}{2 (a n - n + 1) (b n - n + 1)} \left( (a-1)(b-1) n^2 +a+(b-2) n+1 \right. \\
			& \left. + \sqrt{ \big( (n-1)(b n - n + 1) - a \left((b-1) n^2 + 1\right) \big)^2 
			- 4 a (b-1)(n-1) (a n - n + 1) (b n- n + 1)} \right) \,.
	\end{aligned}
\label{eq:pstarNN}
\end{equation}

With these preliminary expressions, we can calculate the critical error probability.  From Eq.~\eqref{eq:bcdefnn}, the left-hand side is
\begin{align}
  P(x_{t+1} = 1|y_{t+1} = 2, y^t = 1^t) &= \frac{ P(y_{t+1} = 2|x_{t+1} 
  	= 1, \cancel{y^t = 1^t}) P(x_{t+1} = 1|y^t = 1^t) }{ P(y_{t+1} = 2|y^t = 1^t) } \notag \\
  &=  \frac{1}{Z_{t+1}}P(y_{t+1} = 2|x_{t+1} = 1) P(x_{t+1} = 1|y^t = 1^t) \,.
\end{align}
The right-hand side is expanded in the same way.  Writing out the individual terms of the equation leads to
\begin{subequations}
\begin{align}
	\lim_{t \to \infty} P(x_{t+1} = 1|y^t = 1^t) 
		&= \lim_{t \to \infty} \sum_{x_t} P(x_{t+1} = 1|x_{t}) P(x_{t}|y^{t} = 1^t)\notag \\
		&= \left( 1 - a \right) p^{*} + (n-1) \frac{a}{n-1} \frac{1-p^{*}}{n-1} \notag \\[5pt]
		&= \left( 1 - a \right) p^{*} + a \frac{1-p^{*}}{n-1} \,,  \\
%
\text{and} \qquad Z_{t+1} &= P(y_{t+1} = 2|y^t = 1^t) \notag \\
	&= \sum_{x_{t+1}} P(y_{t+1} = 2|x_{t+1}) P(x_{t+1}|y^t = 1^t) \,.
\end{align}
\end{subequations}
Plugging all these terms into Eq.~\eqref{eq:bcdefnn} and substituting $p^{*}$ from Eq.~\eqref{eq:pstarNN}, we solve for the critical error probability $b_{\rm c}$ as a function of $a$ and $n$ and find Eq.~\eqref{eq:threshold_nstates}.


\end{widetext}

\begin{acknowledgments}

We thank Malcolm Kennett for suggestions concerning the Ising-model map and Rapha\"el Ch\'etrite for comments on the manuscript.  This work was funded by NSERC (Canada).

\end{acknowledgments}

\end{document}